\documentclass[
reprint,
amsmath,
amssymb,
aps,
showkeys,
]{revtex4-2}

\usepackage{graphicx}
\usepackage{dcolumn}
\usepackage{bm}
\usepackage{natbib}
\usepackage{upgreek}

\usepackage{xcolor}

\begin{document}


\title{Near-lossless method for generating thermal photon-bunched light}

\author{Xi Jie Yeo$^{1}$}
\author{Darren Ming Zhi Koh$^{1}$}
\author{Justin Yu Xiang Peh$^{1}$}
\author{Christian Kurtsiefer$^{1,2}$}
\author{Peng Kian Tan$^{1}$}
\email{cqttpk@nus.edu.sg}
\affiliation{$^{1}$Centre for Quantum Technologies, 3 Science Drive 2, Singapore 117543}
\affiliation{$^{2}$Department of Physics, National University of Singapore, 2 Science Drive 3, Singapore, 117551}

\date{\today}

\begin{abstract}
Thermal light sources exhibiting photon bunching have been suggested for sensing applications that exploit timing correlations of stationary light, including range finding, clock synchronization, and non-line-of-sight imaging.
However, these proposals have remained unrealized in practice because available sources of photon bunching either possess coherence times too short to be timing resolved by photodetectors, or produce brightness levels too low to tolerate realistic return losses.
In this work, we demonstrate a low-loss method for generating photon bunching with a conversion efficiency nearly 9 orders of magnitude higher than that achieved by many other bunching processes.
\end{abstract}


\maketitle
\section{Introduction}
Many sensing methods with non-classical light generated  by parametric
conversion mostly rely on the tight temporal correlation of photon
pairs~\cite{Clark:21}. However, a significant temporal correlation between
photodetection times can also be found in thermal light~\cite{loudon:00}.
\begin{figure}
	\centering\includegraphics[width=\columnwidth]{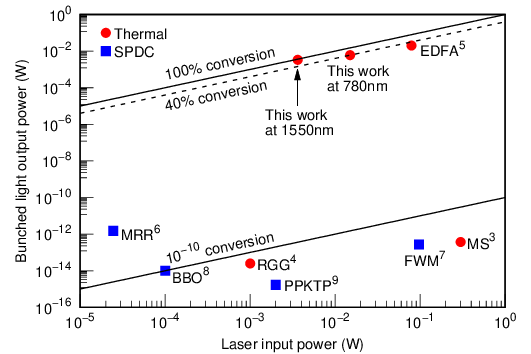}
	\caption{\label{fig:conversion}
		Power relations of various methods and for generating photon-bunched light
		from laser light: MS \cite{dravins:15} -- liquid suspension of microspheres,
		RGG \cite{zhu:12} -- rotating ground glass, EDFA \cite{janassek:18} --
		Erbium-doped fiber amplifier. Methods based on parametric conversion: MRR
		\cite{steiner:21} -- cavity enhanced via microring resonator, FWM
		\cite{england:19} -- four-wave mixing, BBO \cite{lohrmann:18} -- downconversion in $\beta$-Barium Borate crystal, PPKTP \cite{jeong:16} -- downconversion in periodically polled Potassium Titanyl Phosphate crystal}
\end{figure}
Blackbody radiation provides a natural source of thermal photon bunching, with Sunlight filtered at 546\,nm exhibiting a spectral density comparable to that of a low-pressure Mercury vapor discharge lamp, on the order of $10^{7}$ photoevents per second per nanometer per mode.
Much brighter sources of photon-bunched light are based on conversion of
coherent laser light \cite{dravins:15, zhu:12, janassek:18, steiner:21,
	england:19, lohrmann:18, jeong:16} (see Fig.~\ref{fig:conversion}).
Already established methods, commonly referred to as sources of `pseudothermal
light', rely on scattering coherent laser light from random phase-dispersive media, such as a rotating ground glass, or a liquid suspension of microspheres in Brownian motion.

\begin{figure}
	\centering\includegraphics[width=\columnwidth]{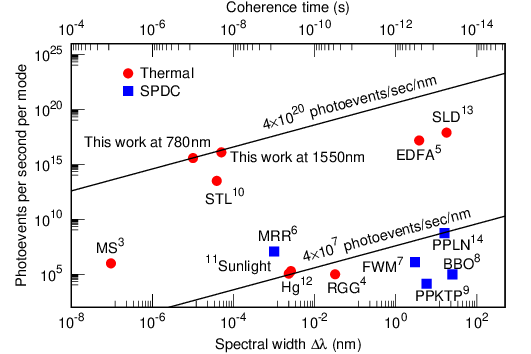}
	\caption{\label{fig:comparison} 
		Spectral densities of stationary light sources exhibiting photon-bunching
		correlations. Apart from the demonstrations referenced in
		Fig.~\ref{fig:conversion}, other demonstrations shown are: STL \cite{pk:23} --
		subthreshold laser diode, \cite{pk:16} -- filtered Sunlight, Hg \cite{pk:14}
		-- Mercury discharge lamp, SLD \cite{rahman:20} -- superluminescent diode,
		PPLN \cite{zhang:15} -- parametric down conversion in periodically poled Lithium Niobate.}
\end{figure}

The scattering processes in these methods introduce speckle patterns.
These scattering-induced fluctuations lead to highly inefficient coupling to
single spatial optical modes, reducing the intensity of bunched light by approximately 10 orders of magnitude relative to the input laser.
Photon bunched light can also be prepared through spontaneous parametric down-conversion (SPDC) processes which are similarly inefficient.

Erbium-doped fiber amplifiers (EDFA) operating below their lasing threshold
can reach a relatively high conversion efficiency, transforming about 25\% of
the 980\,nm pump laser into amplified spontaneous emission centered around
1550\,nm. However, the resulting output thermal light is spectrally broad,
spanning several tens of nanometers, with a correspondingly short
sub-picosecond coherence time that cannot be readily resolved by conventional
photodetectors for observing photon bunching.
Moreover, the accessible conversion wavelengths are inherently restricted by
the amplification process. A chart of the relationship between spectral
bandwidth and photon output per optical mode for demonstrated photon-bunched
light sources is shown in Fig.~\ref{fig:comparison}.

Here, we demonstrate a wavelength-independent technique for generating
spectrally narrowband photon-bunched light from laser light.
In principle, the process can approach a near-unity conversion efficiency, as
it is free from the fundamental losses typically associated with mode mismatch
associated with scattering processes.

\section{Idea}
\begin{figure}
	\centering\includegraphics[width=\columnwidth]{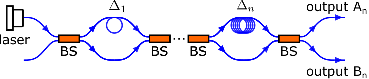}
	\caption{\label{fig:setup} 
		Scheme to efficiently generate thermal photon-bunched light from coherent laser light by using a cascaded series of asymmetric Mach-Zehnder fiber interferometers.
		The input field is first split by a beamsplitter (BS), with one arm acquiring a propagation delay $\Delta_{1}$ before recombination at a subsequent beamsplitter with the non-delayed arm.
		The interferometric loop can be repeated $n$ times, with each stage
		introducing progressively longer fiber delays $\Delta_{n}$, providing two single-mode outputs ($A_n, B_n$).
	}
\end{figure}
Thermal light can be modeled as an ensemble of phase-independent coherent fields \cite{loudon:00}.
To generate thermal photon-bunched light from laser light, a scheme is
required to transform the laser emission into a micro-ensemble of mutually
phase-independent fields. We rely on the random phase relation of
typical laser light at time differences outside the coherence time of the
laser.  
The conversion can be realized by cascading a series of asymmetric
Mach-Zehnder interferometers, each introducing an additional propagation delay
$\Delta_n$ in one arm (see Fig.~\ref{fig:setup}).
The accumulated delays randomize the relative phases of the component fields without introducing scattering losses.
This scheme can be implemented in single-mode optical fibers, suppressing the
losses from mode mismatch compared to a free space implementation.

The electric field amplitude of the input laser light can be described by
\begin{equation}            
	E(t) = E_0e^{i\left[2\pi f t +\phi(t)\right]}\,,
	\label{eqn:E-field-laser}
\end{equation}
where $E_{0}$ denotes the constant electric field amplitude, $f$ the laser
frequency, and $\phi(t)$ the time-dependent phase fluctuation capturing the
finite coherence time of the laser \cite{mandel:59}.

After the light passes through the first interferometer with an additional
delay of $\Delta_1$ in one arm, the electric field amplitudes (for matching
polarizations at the second beamsplitter) at outputs A$_1$, B$_1$ are
\begin{equation}
	E_{A_1,B_1}(t)=\frac{1}{2}[E(t) \pm E(t-\Delta_1)]\,,
	\label{eqn:first-splitter-output}  
\end{equation}
with the difference between the two outputs caused by the relative $\pi$ phase
shift imparted by any beamsplitter. 

Passing both output fields through the next interferometer with a
corresponding delay $\Delta_2$ produces the following field amplitudes (for
matching polarizations) at outputs $A_2$, $B_2$:
\begin{align}
	E_{A_2,B_2}(t) =& \frac{1}{2\sqrt{2}}[E(t) - E(t-\Delta_1) \mp E(t-\Delta_2) 		\nonumber \\
	&\mp E(t-\Delta_1-\Delta_2)]\,.
\end{align}
These light fields are a superposition of the initial laser field $E(t)$ with delayed copies of itself through a combination of delays $\Delta_1, \Delta_2$.

Repeating this procedure through $n$ cascaded interferometers leads to output fields that are superpositions of the original coherent field and all delayed copies generated by the full set of delays $\{\Delta_{1},\Delta_{2},...,\Delta_{n}\}$,
\begin{align}
	E_{A_n}(t) = &\frac{1}{\sqrt{2}^{\,n+1}} \Big[E(t) + \sum_{j=1}^{n}\alpha_{j}E(t-\Delta_j)\nonumber\\
	&+\sum_{j=1}^{n-1}\sum_{k=j+1}^{n}\alpha_{jk}E(t-\Delta_j-\Delta_k) 
	\nonumber\\
	&+\sum_{j=1}^{n-2}\sum_{k=j+1}^{n-1}\sum_{l=k+1}^{n}\alpha_{jkl}E(t-\Delta_j-\Delta_k-\Delta_l) 
	\nonumber\\
	&+\cdots 
	\nonumber\\
	&+E(t-\Delta_1-\cdots-\Delta_n)\Big]\,,
\end{align}
with all coefficients $\alpha \in \{1,-1\}$. A similar expansion holds for
$E_{B_n}$. The resulting output fields are superpositions of the initial laser light field $E(t)$ together with the $2^n-1$ delayed copies of itself that went through different combinations of delays $\Delta_1$ to $\Delta_n$.

To ensure phase independence between the various delayed copies of the initial light field $E(t)$, the delays $\Delta_n$ in each interferometric loop $n$ are chosen to satisfy
\begin{equation}
	\Delta_{n}-\Delta_{n-1} \geq \tau_c, 
	\label{eqn:delta}
\end{equation}
such that each successive delay exceeds the laser coherence time $\tau_{c}$.

According to the Wiener-Khinchin Theorem, the autocorrelation function of a stationary signal is given by the Fourier transform of its power spectral density.
A laser with a Lorentzian spectral line shape \cite{loudon:00} therefore exhibits a first-order coherence function $g^{(1)}(\tau)$ proportional to $e^{-\tau/\tau_{c}}$, whereby $1/\tau_{c}$ is the corresponding laser linewidth.

The output is therefore an ensemble of $2^{n}$ phase-independent coherent fields.
As a result, the light exhibits thermal photon bunching characterized by the second-order coherence function:
\begin{equation}
	g^{(2)}(\tau) - 1 = \bigg(1 - \frac{1}{2^{n}} \bigg)e^{-2|\tau|/\tau_{c}}.
	\label{eqn:g2}
\end{equation}
The photon bunching amplitude increases with the number of interferometric
loops $n$, converging to $g^{(2)}(0)-1=1$ for an infinite number of
phase-independent light fields, matching the correlation of an ideal thermal
light source.

\section{Experimental implementation}


To demonstrate the idea, we inject light from a distributed feedback laser
providing linearly polarized light with a center wavelength of 780\,nm into a single spatial mode optical fiber and couple this to a fiber beamsplitter
separating it into two modes. The first asymmetric Mach-Zehnder
interferometer passes light through an additional optical fiber
to introduce a propagation delay $\Delta_{1}$ in one arm with an additional
optical fiber of $l_1=100$\,m length (corresponding to $\Delta_{1}\approx 495$\,ns),
exceeding the laser coherence time $\tau_{c}$, and another fiber beam splitter
closing the interferometer.

A paddle polarization controller (not shown in Fig.~\ref{fig:setup}) ensures the simple form of adding amplitudes in
Eq.~(\ref{eqn:first-splitter-output}).
Additionally, variable attenuators are inserted in the short leg of the interferometer to compensate for losses in the delay fibers, thereby balancing light intensities just prior to mode-overlap at the subsequent beamsplitters. 
These parameters are optimized to increase the interferometric visibility; the
maximum of the photon bunching signature $g^{(2)}(\tau=0)$ is taken as
an optimization criterion for polarization and intensity balancing control.

To chain the delay combination, more asymmetric Mach-Zehnder interferometers are
added, as well as polarization and intensity balance components. The delay
fiber lengths ($l_2=186$\,m, $l_3=500$\,m) are chosen to introduce
propagation delays $\Delta_{n}$ in each loop that closely satisfy the condition Eq.~(\ref{eqn:delta}).


To perform the $g^{(2)}(\tau)$ timing correlation measurements, light from
either output $A_n$ or $B_n$ is  collected by a pair of actively quenched
Silicon avalanche photodetectors. The resulting photodetection events are
timestamped with a time resolution of 2\,ns. 

\section{Results}
\begin{figure}
	\centering
	\includegraphics[width=\columnwidth]{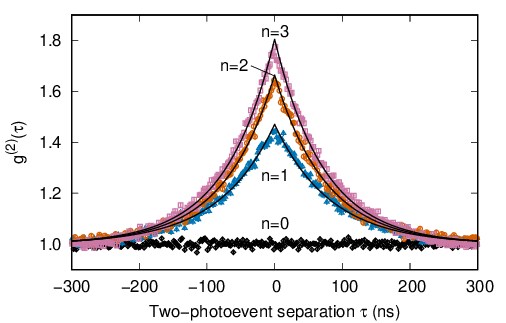}
	\caption{\label{fig:g2}
		Photon bunching signatures from a laser light (center wavelength
		$\lambda=780$\,nm)  passing through a number $n$ of
		asymmetric Mach-Zehnder interferometers.
		A numerical fit to a double-exponential decay reaches a
		peak value of the second order correlation function at time delay $\tau=0$. We
		find for $n=0$ (no interferometer): $g^{(2)}(\tau)=1$, 
		for $n=1$: $g^{(2)}(0)=1.471\pm0.003$, 
		for $n=2$: $g^{(2)}(0)=1.665\pm0.003$,  
		and for $n=3$: $g^{(2)}(0)=1.805\pm0.004$. 
		The coherence times extracted for all traces is around $\tau_c\approx135$\,ns
		(see text for details). } 
\end{figure}

To show the photon bunching signature, the temporal correlation functions
$g^{(2)}(\tau)$ for a different number $n$ interferometers were numerically
extracted from the stream of detector time stamps, and are shown in
Fig.~\ref{fig:g2}.

As a reference, a flat $g^{(2)}(\tau)=1$ is observed for $n=0$ (i.e., no
interferometer) which is consistent for the un-bunched Poisson statistics
expected from a laser emitting coherent light.
The resultant photon bunching peaks for $n>0$ were fitted to two-sided
exponential decays (as expected for Lorentzian line shapes of the laser
light), using a model function $g^{(2)}(\tau) = 1 + b\cdot
e^{-2|\tau|/\tau_c}$.
For $n=1$, we obtain $g^{(2)}(0)=1.471\pm0.003$ with a coherence time of
$\tau_{c}=135.6\pm0.3$\,ns. For $n=2$, we obtain $g^{(2)}(0)=1.665\pm0.003$
and $\tau_{c}=134.8\pm0.2$\,ns, for $n=3$, $g^{(2)}(0)=1.805\pm0.004$ and
$\tau_{c}=135.2\pm0.2$\,ns, respectively.
This suggests that the correlation function converges towards
$g^{(2)}(\tau=0)=2$ with increasing number $n$ of interferometers (providing
an increasing number of phase-independent effective light sources with the
same spectrum).
This behavior closely approximates the statistics of an ideal thermal light source with photon bunching characteristics  $g^{(2)}(\tau) = 1 + e^{-2|\tau|/\tau_c}$.

\begin{figure}
	\includegraphics[width=\columnwidth]{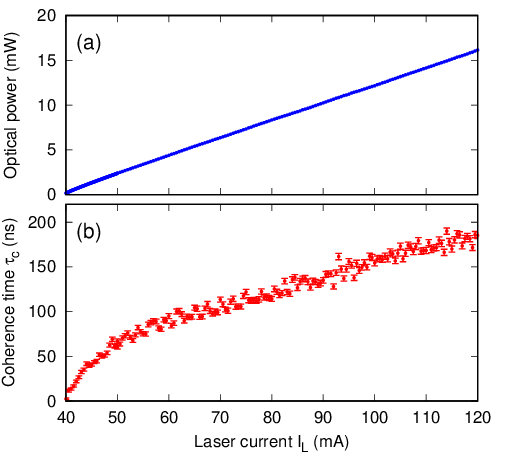}
	\caption{\label{fig:power_coherence_time}
		Tunable (a) output photon bunching power and (b) coherence time $\tau_c$ by adjusting the laser injection current $I_{L}$.}
\end{figure}

The coherence time $\tau_c$ of the laser used was long which conveniently allowed us
to resolve the exponential decay of the correlation function from the thermal
bunching peak. We can also vary the coherence time by adjusting
the injection current to the laser diode. In
Fig.~\ref{fig:power_coherence_time}, we show the dependency of the laser power
and coherence time $\tau_c$.
For injection currents slightly above the lasing threshold (around 40\,mA), the
laser emits very little power and has a short coherence time. With
increasing injection current, the output power of the laser increases, as well
as the coherence time extracted from the decay of $g^{(2)}(\tau)$. Therefore, 
the presented technique also allows control over the temporal coherence
time, complementing the control over the photon bunching amplitude
$g^{(2)}(\tau=0)$ with the number $n$ of asymmetric Mach-Zehnder interferometers.

For $n=1$, we observe a conversion efficiency from laser light to
photon-bunched light at both output ports together of 40\%. This loss is
mostly contributed by the high losses in our delay fiber at 780\,nm wavelength,
and the corresponding attenuation to balance the intensity in both
interferometer arms. A significantly higher conversion efficiency can be
obtained when moving to a wavelength where optical fibers have a lower loss.

\begin{figure}
	\centering\includegraphics[width=\columnwidth]{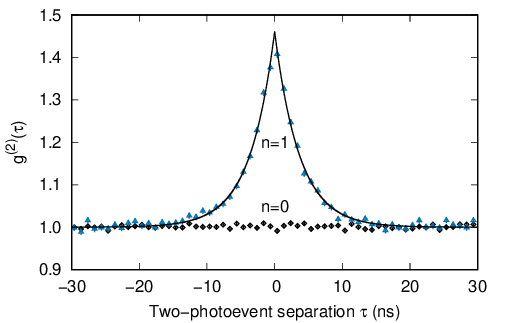}
	\caption{\label{fig:1550nm}
		Photon bunching signature from laser light  ($\lambda=1550$\,nm)
		passing through $n=1$ interferometer. Numerical fitting gives a
		peak $g^{(2)}(\tau=0)=1.461\pm0.006$ with a coherence time of
		$\tau_{c}=7.4\pm0.2$\,ns.}
\end{figure}

With a similar experimental setup at the telecomm-wavelength of $1550\,$nm, we
observe a combined power from both output ports of $3.36\,$mW from an initial
input power $3.63\,$mW, corresponding to a conversion efficiency of 92.5\%. A clear photon bunching peak can also been seen with $g^{(2)}(\tau=0)=1.461\pm0.006$ with a coherence time of $\tau_{c}=7.4\pm0.2$\,ns as shown in Fig.~\ref{fig:1550nm}.\vspace{1mm}


\section{Conclusion and Outlook}
We demonstrated a technique to generate thermal photon bunching that is about
9 orders of magnitude more efficient than many other thermal or parametric
down conversion processes, showing a photon pair timing correlation that could
have applications that rely on time correlations between two photons with an
intrinsic randomness of single-photon timings. Moreover, this method also
allows for control over the peak photon bunching signature and temporal
coherence. If optical wavelengths are used that lead to negligible losses
in the asymmetric interferometers, an almost unit conversion efficiency of
laser light to photon-bunched light can be reached, providing an unprecedented
brightness of photon-bunched light. 
This efficiency and tunability enable the sensing system to tolerate high return losses in realistic environments, thereby opening a pathway toward practical sensing implementations such as range finding, clock synchronization, and non-line-of-sight imaging.

\section{Acknowledgments}
This research is supported by the Quantum Engineering Programme through NRF2021-QEP2-03-P02, the Ministry of Education, and the National Research Foundation, Prime Minister's Office, Singapore.

\vfill
%

\end{document}